\documentclass[1p,preprint,10pt]{elsarticle}
\usepackage{amsmath}
\usepackage{amssymb}
\usepackage{amsmath}
\usepackage{textcomp}
\usepackage{color}
\usepackage{framed}
\usepackage[usenames,dvipsnames]{xcolor}
\usepackage{hyperref}
\hypersetup{colorlinks=true,breaklinks=true,linkcolor=blue,urlcolor=blue,citecolor=blue}
\usepackage{adjustbox}
\usepackage{tikz}
\usepackage{float}
\usetikzlibrary{shapes,arrows,positioning,fit}
\usetikzlibrary{calc}

\renewcommand{\d}[1]{\tilde #1}
\newcommand{\ndag}{{\dagger}}

\newcommand{\up}{\ensuremath{\uparrow}}
\newcommand{\dn}{\ensuremath{\downarrow}}

\newcommand{\ket}[1]{| #1 \rangle}

\renewcommand{\d}{\ensuremath{\mathrm{d}}}

\newcommand{\Ttau}{{\ensuremath{\cal T}}}

\newcommand{\moy}[1]{\ensuremath{\left\langle #1 \right\rangle}}

\tikzstyle{decision} = [diamond, draw, fill=green!20,
    text width=4.5em, text badly centered, node distance=3cm, inner sep=0pt]
\tikzstyle{block} = [rectangle, draw, fill=blue!20,
    text width=5em, text centered, rounded corners, minimum height=2em]
\tikzstyle{terminal} = [ellipse, draw, fill=red!20,
    text width=5em, text centered, minimum height=2em]
\tikzstyle{line} = [draw, -latex']

\makeatletter
\tikzset{
  fitting node/.style={
    inner sep=0pt,
    fill=none,
    draw=none,
    reset transform,
    fit={(\pgf@pathminx,\pgf@pathminy) (\pgf@pathmaxx,\pgf@pathmaxy)}
  },
  reset transform/.code={\pgftransformreset}
}
\makeatother

\journal{Computer Physics Communications}

\usepackage{lineno}

\definecolor{darkblue}{rgb}{0,0,.6}
\definecolor{darkred}{rgb}{.6,0,0}
\definecolor{darkgreen}{rgb}{0,.6,0}
\definecolor{red}{rgb}{.98,0,0}

\def\ssmall{\fontsize{8pt}{2pt}\selectfont}

\usepackage{listings}
\usepackage{caption}

\DeclareCaptionFormat{listing}{\rule{\dimexpr\textwidth+17pt\relax}{0.4pt}\par\vskip1pt#1#2#3}
\lstnewenvironment{cpplisting}[1][]
  {\lstset{language=C++,numbers=right,#1}}{}

\lstnewenvironment{pylisting}[1][]
  {\lstset{language=Python,numbers=right,#1}}{}

\newcommand\py[1]{\lstinline[language=Python]{#1}}

\newcounter{bla}

\begin{document}

\begin{frontmatter}

\title{TRIQS/CTHYB: A Continuous-Time Quantum Monte Carlo Hybridization Expansion Solver for Quantum Impurity Problems}

\author[X]{Priyanka Seth\corref{author}} \ead{priyanka.seth@polytechnique.edu}
\author[Hamburg]{Igor Krivenko} \ead{ikrivenk@physnet.uni-hamburg.de}
\author[X]{Michel Ferrero} \ead{michel.ferrero@polytechnique.edu}
\author[CEA]{Olivier Parcollet} \ead{olivier.parcollet@cea.fr}

\cortext[author] {Corresponding author.}

\address[X]{\'{E}cole Polytechnique, CNRS, 91128 Palaiseau Cedex, France}
\address[CEA]{Institut de Physique Th\'eorique (IPhT), CEA, CNRS, 91191 Gif-sur-Yvette, France}
\address[Hamburg]{I. Institut f\"ur Theoretische Physik, Uni. Hamburg, Jungiusstra\ss e 920355 Hamburg, Germany}

\begin{abstract}

We present \py{TRIQS/CTHYB}, a state-of-the art open-source implementation of
the continuous-time hybridisation expansion quantum impurity solver of the
TRIQS package. This code is mainly designed to be used with the TRIQS library
in order to solve the self-consistent quantum impurity problem in a multi-orbital
dynamical mean field theory approach to strongly-correlated electrons, in
particular in the context of realistic calculations. It is implemented in C++
for efficiency and is provided with a high-level Python interface.
The code is ships with a new partitioning algorithm that
divides the local Hilbert space without any user knowledge of the
symmetries and quantum numbers of the Hamiltonian.
Furthermore, we implement higher-order configuration moves and show that such
moves are necessary to ensure ergodicity of the Monte Carlo in common
Hamiltonians even without symmetry-breaking.

\end{abstract}

\end{frontmatter}



\noindent {\bf PROGRAM SUMMARY}

\begin{small}
\noindent
{\em Program Title:} TRIQS/CTHYB                                    \\
{\em Project homepage:} http://ipht.cea.fr/triqs/applications/cthyb \\
{\em Catalogue identifier:} --                                      \\
{\em Journal Reference:} --                                         \\
{\em Operating system:} Unix, Linux, OSX \\
{\em Programming language:} \verb*#C++#/\verb*#Python#\\
{\em Computers:} 
  any architecture with suitable compilers including PCs and clusters \\
{\em RAM:} Highly problem-dependent \\
{\em Distribution format:} GitHub, downloadable as zip \\
{\em Licensing provisions:} GNU General Public License (GPLv3)\\
{\em Classification:} 4.4, 4.8, 4.12, 6.5, 7.3, 20  
{\em External routines/libraries:} 
  \verb#TRIQS#,
  \verb#cmake#.\\
{\em Nature of problem:}\\
Accurate solvers for quantum impurity problems are needed in condensed matter theory.\\
{\em Solution method:}\\
We present an efficient \verb#C++#/\verb#Python# open-source implementation of
a continuous-time hybridization expansion solver.
\end{small}
\section{Introduction}
\label{sec:introduction}

Strongly-correlated quantum systems are a central challenge for theoretical
condensed matter physics with a wide range of remarkable phenomena such as
metal-insulator transitions, high-temperature superconductivity, and magnetism.
In the last two decades, tremendous progress has been made in the field of
algorithms for the quantum many-body problem, both in refining existing
techniques and in developing new systematic approximations and algorithms.
Among these methods is dynamical mean-field theory (DMFT) 
\cite{georges_dynamical_1996,kotliar_electronic_2006} and its cluster
\cite{maier_quantum_2005} or diagrammatic extensions
\cite{toschi_dynamical_2007,rubtsov_dual_2008,rubtsov_dual_2012}. DMFT methods
can also be combined with more traditional electronic structure methods
such as density functional theory leading to \emph{ab initio} realistic
computational techniques for strongly-correlated materials
\cite{kotliar_electronic_2006}.

Quantum impurity models play a central role in DMFT methods. From the
computational point of view, they are normally the bottleneck. The study of
modern algorithms for these problems, and their implementation is therefore of
great importance in the field. In the last decade, the continuous-time quantum
Monte Carlo methods have emerged and become well-established, starting with the
work of A.~Rubtsov {\it et al.}~\cite{rubtsov_continuous-time_2005}, P.~Werner
{\it et al.}~\cite{werner_continuous-time_2006,werner_hybridization_2006,
gull_continuous-time_2011}, 
and E.~Gull {\it et al.}~\cite{gull_continuous-time_2008,gull_ctaux_2008}. 
In the continuous-time quantum Monte Carlo algorithm based on the hybridization
expansion (CT-HYB) \cite{werner_continuous-time_2006,gull_continuous-time_2011},
the Monte Carlo performs a systematic expansion in the hybridization
of the impurity to the bath. In the last years, several improvements have been
proposed to optimise such algorithms
\cite{haule_quantum_2007,gull_continuous-time_2008,augustinsky_improved_2013,semon_lazy_2014},
which in particular have greatly improved our capability to solve multi-orbital
models (e.g., 3 or 5 bands) that are crucial in {\it realistic} DMFT
calculations.

In this paper, we present an implementation of the CT-HYB algorithm with
state-of-the-art improvements and optimisations. This code, \py{TRIQS/CTHYB}, is
part of the TRIQS family, and is based on the TRIQS library \cite{triqs_2015}.
It is released under the Free Software GPLv3 license. The algorithm can {\it a
priori} handle any type of quantum impurity models and any interaction form.
The code implements an optimisation of the atomic trace computation using a
balanced left-leaning red-black tree, following the pioneering work of E.~Gull
\cite{gull_continuous-time_2008}, in combination with some controlled
truncation of the trace computation \cite{semon_lazy_2014}
and `quick-abandon' procedure \cite{yee_towards_2012}. In
Sec.~\ref{sec:tree}, we will indeed show explicitly that Gull's balanced tree
algorithm drastically improves the {\it scaling} of the code for a five band
model, in which the computation time increases only very mildly when temperature
decreases.

In addition, we present and implement a new algorithm to divide the
local Hilbert space of the quantum impurity to accelerate matrix products of
the trace computation, without the need to specify its symmetry or a set of
quantum number operators. This is particularly convenient in complex,
multi-orbital contexts, where the set of quantum numbers may not be obvious.
For example, in \cite{parragh_conserved_2012} an additional set of quantum
numbers was discovered for the Kanamori Hamiltonian, leading to a significant
efficiency gain. With the present algorithm, these quantum numbers are no
longer needed as user input. In many cases, this algorithm will lead to a more
thorough Hilbert space decomposition than using only the well-known quantum
numbers.

Let us note that this code is a replacement for the previous implementation of
the \py{CTHYB} package. Indeed all the implemented improvements 
give strictly the same Markov chain as our previous
implementation, so it is the {\it same} Monte Carlo algorithm, but is much
faster in practice. We discuss the speed improvement in more detail in
Sec.~\ref{sec:tree}.

For systems with broken symmetry, it was shown that higher-order configuration
moves were crucial to ensure ergodicity of the algorithm and hence give correct
results. We implement such moves in the quantum Monte Carlo and find that such
moves are required to ensure ergodicity even in the absence of
symmetry-breaking \cite{semon_ergodicity_2014}. An example of such a case is
given in Sec.~\ref{sec:doublemoves}.

This paper is organized as follows. In Sec.~\ref{sec:usage}, we first show in a
simple example how to use the solver from its high-level Python interface. In
Sec.~\ref{sec:cthyb}, we give a general overview of the CT-HYB formalism.
We discuss partitioning schemes in Sec.~\ref{sec:partitioning}.
We briefly describe the tree-based optimisation of the dynamical trace computation in
Sec.~\ref{sec:tree}. In Sec.~\ref{sec:doublemoves}, we discuss ergodicity of
the algorithm and the need for higher-order configuration updates. We give
detail on how to obtain and install the package in
Sec.~\ref{sec:starting}, and finally conclude in Sec.~\ref{sec:summary}.

\section{Usage}
\label{sec:usage}

The \py{CTHYB} application is built on the TRIQS library \cite{triqs_2015},
which must be pre-installed. The interface is very simple. The user
initialises the \py{Solver} object, sets the input quantities, and finally calls the
solver. The resulting output quantities can thereafter be analysed and
manipulated by means of the TRIQS library.\footnote{The interface described
above is similar to that of of the simpler CT-INT algorithm presented as an
example in Ref.~\citealp{triqs_2015}.}

Along the lines of the TRIQS library, the core of the \py{CTHYB} solver presented
here is written in C++ for efficiency, with a Python wrapper for ease of use.
Hence the solver can be used directly from C++, in a Python script or
interactively in an IPython notebook.

Below we show an example of a Python script implementing a DMFT loop for a 
five-band system on the Bethe lattice with fully rotationally-invariant
interactions.

\begin{lstlisting}[language=Python,
		 caption=Example script to run DMFT on a five-band model with
                         fully rotationally-invariant Slater interactions,
                 label=5bandslater,numbers=left]
from pytriqs.gf.local import *
from pytriqs.archive import HDFArchive
from pytriqs.applications.impurity_solvers.cthyb import Solver
import pytriqs.operators.util as op
import pytriqs.utility.mpi as mpi

# General parameters
filename = 'slater_5_band.h5'                # Name of file to save data to
beta = 100.0                                 # Inverse temperature
l = 2                                        # Angular momentum
n_orbs = 2*l + 1                             # Number of orbitals
U = 5.0                                      # Screened Coulomb interaction
J = 0.1                                      # Hund's coupling
half_bandwidth = 1.0                         # Half bandwidth
mu = (U/2.0)*((2.0*n_orbs)-1.0) \
     - (5.0*J/2.0)*(n_orbs-1.0)              # Half-filling chemical potential
spin_names = ['up','down']                   # Outer (non-hybridizing) blocks
orb_names = ['%s'%i for i in range(n_orbs)]  # Orbital indices
off_diag = True                              # Include orbital off-diagonal elements
n_loop = 5                                   # Number of DMFT self-consistency loops

# Solver parameters
p = {}
p["n_warmup_cycles"] = 10000     # Number of warmup cycles to equilibrate
p["n_cycles"] = 10000            # Number of measuresments
p["length_cycle"] = 100          # Number of MC steps between consecutive measures
p["move_double"] = True          # Use four-operator moves

# Block structure of Green's functions
# gf_struct = {'up':[0,1,2,3,4], 'down':[0,1,2,3,4]}
# This can be computed using the TRIQS function as follows:
gf_struct = op.set_operator_structure(spin_names,orb_names,off_diag=off_diag) 

# Construct the 4-index U matrix U_{ijkl}
# The spherically-symmetric U matrix is parametrised by the radial integrals
# F_0, F_2, F_4, which are related to U and J. We use the functions provided
# in the TRIQS library to construct this easily:
U_mat = op.U_matrix(l=l, U_int=U, J_hund=J, basis='spherical')

# Set the interacting part of the local Hamiltonian
# Here we use the full rotationally-invariant interaction parametrised 
# by the 4-index tensor U_{ijkl}.
# The TRIQS library provides a function to build this Hamiltonian from the U tensor:
H = op.h_int_slater(spin_names,orb_names,U_mat,off_diag=off_diag)

# Construct the solver
S = Solver(beta=beta, gf_struct=gf_struct)

# Set the hybridization function and G0_iw for the Bethe lattice
delta_iw = GfImFreq(indices=orb_names, beta=beta)
delta_iw << (half_bandwidth/2.0)**2 * SemiCircular(half_bandwidth)
for name, g0 in S.G0_iw: g0 << inverse(iOmega_n + mu - delta_iw)

# Now start the DMFT loops
for i_loop in range(n_loop):

  # Determine the new Weiss field G0_iw self-consistently
  if i_loop > 0: 
    g_iw = GfImFreq(indices=orb_names, beta=beta)
    # Impose paramagnetism
    for name, g in S.G_tau: g_iw << g_iw + (0.5/n_orbs)*Fourier(g)
    # Compute S.G0_iw with the self-consistency condition
    for name, g0 in S.G0_iw: 
        g0 << inverse(iOmega_n + mu - (half_bandwidth/2.0)**2 * g_iw )

  # Solve the impurity problem for the given interacting Hamiltonian and set of parameters
  S.solve(h_int=H, **p)

  # Save quantities of interest on the master node to an h5 archive
  if mpi.is_master_node():
      with HDFArchive(filename,'a') as Results:
          Results['G0_iw-%s'%i_loop] = S.G0_iw
          Results['G_tau-%s'%i_loop] = S.G_tau
\end{lstlisting}

\subsection{Input parameters}

The basic usage of the solver involves three steps: first, the construction of
a \py{Solver} object, second, the setting of the input Weiss field \py{G0_iw},
and finally, the invoking of the \py{solve()} method. We describe below the
necessary parameters for each of these steps:
\begin{itemize}

  \item \py{S = Solver(beta, gf_struct)} (line 47 in Listing \ref{5bandslater}): 
        Here we construct a \py{Solver} instance called \py{S} with the following parameters:
  \begin{itemize}
     \item \py{beta}:
        The inverse temperature $\beta$.
     \item \py{gf_struct}:
	The structure of the block Green's function (\py{BlockGf} class, see
        TRIQS library documentation) given as a Python dictionary \py{\{block:inner\}},
        where \py{block} is a string and \py{inner} is a Python list of indices within
        \py{block}. 
  \end{itemize}
	This ensures that all quantities within Solver are correctly initialised,
        and in particular that the block structure of all Green's function objects is consistent.

  \item \py{S.G0_iw << ...} (line 52 in Listing \ref{5bandslater}): 
        Here we assign the initial Weiss field $G_0$ as given by Eq.~\ref{eq:G0}.

  \item \py{S.solve(**params)} (line 67 in Listing \ref{5bandslater}):
	Here we solve the impurity problem for the set of parameters
        \py{params} given as a Python dictionary. Amongst these parameters are:
  \begin{itemize}
     \item \py{h_int}: 
	This is a TRIQS \py{Operator} object (see Sec.~8.7 in
        Ref.~\citealp{triqs_2015}) consisting of the interaction terms in the local
        Hamiltonian (i.e., those containing more than two operators).
        In Listing \ref{5bandslater}, we use the full rotationally-invariant Hamiltonian
        \begin{equation*}
        H_{\rm int} = \frac{1}{2}\sum_{ijkl,\sigma\sigma'} U_{ijkl} c_{i \sigma}^\dagger c_{j \sigma'}^\dagger c_{l \sigma'} c_{k \sigma}.
        \end{equation*}
        See the TRIQS library documentation for the \py{pytriqs.operators.util.U_matrix} and
        \py{pytriqs.operators.util.hamiltonians} modules.
 
     \item \py{n_cycles}: 
	This is the number of quantum Monte Carlo measurements made. In the
        case of an MPI calculation, the total number of measurements will be
        \py{n_cycles} for each core.

  \end{itemize}
  
\end{itemize}

We refer the reader to the online documentation for a full and regularly
updated listing of all other \py{Solver} initialisation and \py{solve()}
parameters.

\subsection{Output data}

After the \py{solve()} method has completed, several output quantities are
available for analysis. These include:
\begin{itemize}

  \item \py{S.G_tau}:
	The local impurity Green's function in imaginary time $G(\tau)$.

  \item \py{S.G_iw}:
	The local impurity Green's function in imaginary (Matsubara)
        frequencies $G(i\omega_n)$. This is computed after the \py{solve()} as the
        Fourier transform of $G(\tau)$ if the parameter \py{perform_post_proc} is
        \py{True}. The high-frequency expansion coefficients are fitted according to
        user-provided parameters.

  \item \py{S.Sigma_iw}:
	The local impurity self-energy in imaginary (Matsubara) frequencies
        $\Sigma(i\omega_n)$.  This is computed from \py{S.G0_iw} and \py{S.G_iw} by
        solving Dyson's equation if the parameter \py{perform_post_proc} is \py{True}.
        The high-frequency expansion coefficients are fitted according to user-provided
        parameters.

  \item \py{S.G_l}:
	The local impurity Green's function accumulated in Legendre polynomials
	as described in Ref.~\citealp{boehnke_orthogonal_2011}. In order to measure
        \py{S.G_l}, the parameter \py{measure_g_l} must be set to \py{True} in the
        \py{solve()} parameters.

\end{itemize}

\section{Hybridization expansion of partition and Green's function}
\label{sec:cthyb}

Here we outline the principles of the continuous-time hybridization expansion formalism 
\cite{werner_continuous-time_2006,werner_hybridization_2006,haule_quantum_2007,gull_continuous-time_2011}
briefly. We refer the interested reader to the aforementioned references for
more detailed derivations.

The partition function of the impurity model is given by
\begin{equation}\label{eqn:app:Z}
Z = \int {\cal D} c^{\dagger} {\cal D} c \exp ( - S ).
\end{equation}
The action $S$ takes the form
\begin{gather}
 S = - \iint_{0}^{\beta } \d \tau \d \tau^\prime
 \sum_{\substack{\alpha,\beta}}c^{\dagger }_{\alpha} (\tau)
G_{0, \alpha\beta}^{-1} (\tau ,\tau^\prime)
c^{ }_{\beta} (\tau^\prime )
 + \int_{0}^{\beta } \d \tau \label{Seff}
H_\text{int},
\end{gather}
where
\begin{gather}
 G^{-1}_{0\alpha\beta}(i \omega_n) = (i \omega_n + \mu) \delta_{\alpha\beta} - h^0_{\alpha\beta} - \Delta_{\alpha\beta}(i \omega_{n}). \label{eq:G0}
\end{gather}
and $ \Delta_{\alpha\beta}(i \omega_{n})$ is defined such that it vanishes at
high frequencies.
Certain symmetries of the action allow the Green's function to be reduced into
a block diagonal form with blocks labelled by a `block' index. The most common
such decomposition is by spins, denoted by $\sigma$. The orbital can then be a
further, `inner', index $a$ within the spin block. The index $\alpha$ then
refers to the pair $(\sigma,a)$.

Let us define $\Ttau$ as time ordering operator, the local Hamiltonian
\begin{equation}
  H_{\mathrm{loc}}=H_{\mathrm{int}} + \sum_{\alpha,\beta} h^0_{\alpha\beta}
                                      c^{\dagger }_{\alpha} c^{ }_{\beta},
\end{equation}
and $M$ such that
$[M^{-1}]_{i j} = \Delta_{\alpha_i, \alpha_j'} (\tau_{i} - \tau_{j}')$.
The partition function is expanded in powers of the hybridization function as
\begin{gather}
Z = \sum_{k\geq 0} 
\int \prod_{i=1}^{k} \d\tau_{i} \d \tau^\prime_{i}
\sum_{\alpha_{i}, \alpha'_i } w(k,\{ \alpha_j,\alpha'_j,\tau_j,\tau^\prime_j\}),
\end{gather}
with the quantum Monte Carlo Markov chain weights given by 
\begin{gather}
w(k,\{ \alpha_j,\alpha'_j,\tau_j,\tau^\prime_j\}) \equiv
\det_{1\leq i,j \leq k}
\bigl [
 M^{-1} 
\bigr]_{i j}
\mathop{\text{Tr}}
\left( \Ttau
e^{-\beta H_\mathrm{loc}} \prod_{i=1}^{k} c^{\dagger}_{\alpha_{i}}(\tau_{i}) c_{\alpha'_{i}}(\tau^\prime_{i})
\right).
\label{eq:mc_weight}
\end{gather}

The partition function $Z$ and the average of any function $f$ over the space
of sampled configurations
$ {\cal C} \equiv (k, \{\alpha_j,\alpha'_j,\tau_j,\tau^\prime_j\})$ 
are then given by 
\begin{align}
  Z &= \sum_{\cal C} w({\cal C}), \\ 
  \moy{ f({\cal C}) }
  &= \frac{1}{Z} \sum_{\cal C} w({\cal C}) f({\cal C}).
\end{align}

Specifically, the imaginary-time Green's function can be measured as
\begin{equation}
  G_{\alpha \beta} (\tau) = \frac{-1}{Z \beta} \sum_{\cal C} |w({\cal C})| \,
    \bigl[ \mathrm{sgn}(w({\cal C})) \, \sum_{ij} M_{ij} \delta(\tau_i - \tau^\prime_j + \tau)
    \delta_{\alpha_i \alpha} \delta_{\alpha_j' \beta} \bigr].
\end{equation}

Alternatively, the Green's function can be accumulated in Legendre polynomials
following the prescription given in Ref.~\citealp{boehnke_orthogonal_2011}.

In CT-HYB algorithms, the computational bottleneck for the multi-orbital cases
generally comes from the trace calculation for $k$ operators in
Eq.~\ref{eq:mc_weight}, which needs to be recomputed for each new
configuration. On the other hand, the determinants
can be efficiently computed even for moderately large $k$.
We reduce this computational cost in two main ways. 

As shown in Ref.~\citealp{haule_quantum_2007}, the trace computation can be
optimised by decomposing the local Hilbert space into smaller subspaces such that
there is a one-to-one mapping between subspaces under the application of a
creation or annihilation operator. As a result, the matrices whose products
enter the trace can be of significantly lower dimensionality. Usually, this
decomposition of the Hilbert space relies on user input of the local symmetries
through quantum numbers of the Hamiltonian. We have devised a novel
algorithm that accomplishes this partitioning {\it automatically}, without any
{\it a priori} knowledge of quantum numbers of the system, giving a more efficient
decomposition. This algorithm is detailed in Sec.~\ref{sec:partitioning}. 

The second amelioration that reduces the computational burden of trace
calculation of each configuration is, as first suggested by Gull
\cite{gull_continuous-time_2008}, the use of a tree structure to cache parts
of the trace calculation that are unchanged between configurations. The
bounding properties of the trace 
\cite{yee_towards_2012,semon_lazy_2014} help further reduce the time
spent in the trace calculation. These improvements are the subject of
Sec.~\ref{sec:tree}.

We also note that this algorithm allows for the treatment of all forms of
interaction, including using a rotationally-invariant impurity Hamiltonian
parametrised by the 4-index interaction matrix $U_{\alpha \beta \gamma \delta}$.
However, for more complex interactions, the sign problem can become a
hindrance. If the interactions are of purely density-density form, the CT-HYB
segment picture \cite{werner_continuous-time_2006} can be used at greatly
reduced computational cost.

\section{Partitioning the local Hilbert space}
\label{sec:partitioning}

The calculation of the dynamical trace boils down to multiplication of matrices.
It is then crucial to find a way to minimise the sizes of the matrices being
multiplied. This goal can be achieved by partitioning the local Hilbert space
${\cal H}$ of dimension $N$ into a direct sum of $K$ subspaces ${\cal H}_k$
each of dimension $0 < N_k \le N$.

We wish to find a permutation of the basis vectors such
that (1) the local Hamiltonian is block-diagonal, and (2) all $c_\alpha$ and
$c_\alpha^\ndag$ operators are block matrices with at most one non-zero block
in each row and column. Such a permutation would group the basis states
belonging to the same subspace ${\cal H}_k$ together.

We implement two strategies to do the partitioning: with user-supplied quantum
numbers and automatically. The latter strategy is more universal and chosen by
default. 

\subsection{Quantum numbers}

This is the traditional approach to partitioning of the Hilbert space
\cite{haule_quantum_2007}.  The user provides a list of integrals of motion
(operators) $Q_1,\ldots,Q_L$ such that the $Q_i$ can be expressed as a function of
the density operators $n_{\alpha}$.
Expectation values of these operators are calculated for each Fock state
$\ket{\psi}$, which gives a combination of quantum numbers associated with the
state:
\[
  \langle\psi|Q_1|\psi\rangle, \ldots, \langle\psi|Q_L|\psi\rangle\ \Rightarrow q_1,\ldots, q_L.
\]
All states sharing the same set of quantum numbers belong to the same subspace.

Condition (1) is obviously fulfilled by the
obtained subspaces, as by definition, the Hamiltonian cannot connect states
with different quantum numbers. 
As each operator $Q_l$ corresponds to a function of the operators $n_{\alpha}$, 
condition (2) is also fulfilled.

In many cases this approach works well, but it requires the quantum numbers and
hence some prior analysis of the local Hamiltonian. It may be difficult to
discover an exhaustive set of integrals of motion if the dimension of the local
Hilbert space is large and the interaction form is complex
\cite{parragh_conserved_2012}.

\subsection{Automatic partitioning}

The automatic partitioning algorithm employs no additional \textit{a priori}
information about the symmetry of Hamiltonian $\hat H$. The only input data are the full set of
basis Fock states and the Hamiltonian itself. The algorithm consists of two
sequential phases. In the first phase, the finest possible partition, which
satisfies condition (1) alone is constructed. In the second phase, this
partition is modified to additionally satisfy condition (2).

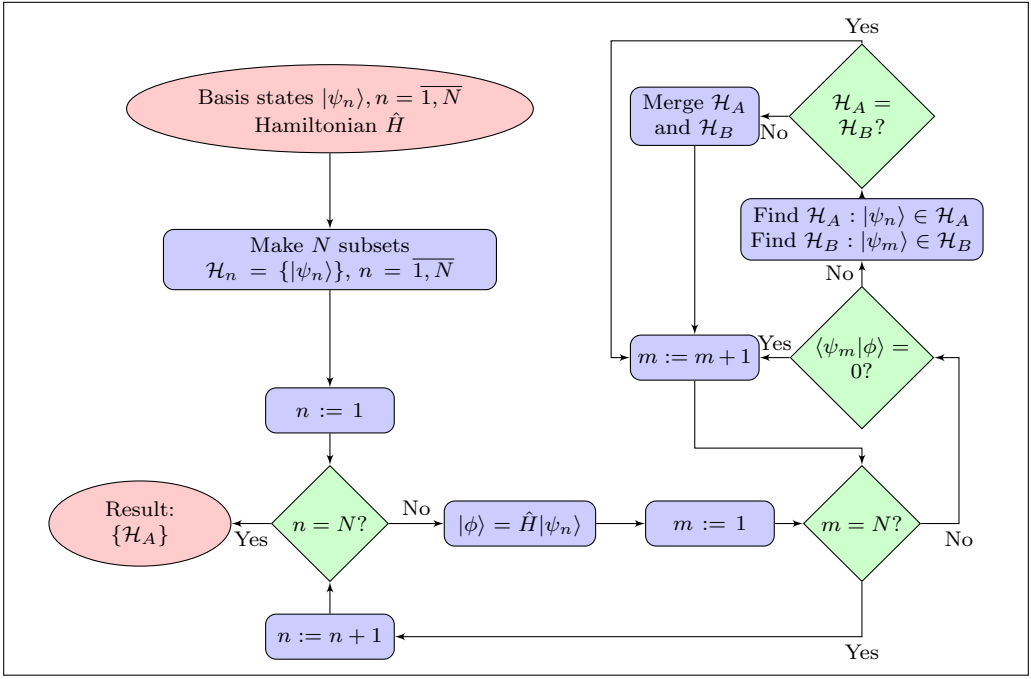
\begin{figure}[H]
    \framebox[\textwidth]{
\begin{tikzpicture}[auto]
font issue=\footnotesize
    \node [terminal, text width=12em] (input) {Basis states $|\psi_n\rangle, n=\overline{1,N}$\par Hamiltonian $\hat H$};
    \node [block, below of=input, text width=14em, node distance=2.0cm] (make_subsets) {Make $N$ subsets ${\cal H}_n=\{|\psi_n\rangle\}$, $n=\overline{1,N}$ };
    \node [block, below of=make_subsets, node distance=2.0cm] (n_loop_enter) {$n:=1$};
    \node [decision, below of=n_loop_enter, node distance=1.5cm] (n_loop_cond) {$n=N$?};
    \node [block, below of=n_loop_cond, node distance=1.5cm] (n_loop_inc) {$n:=n+1$};
    \node [terminal, left of=n_loop_cond, node distance=2.5cm] (result) {Result: $\{{\cal H}_A\}$};
    \node [block, right of=n_loop_cond, node distance=2.5cm, text width=6em] (apply) {$|\phi\rangle = \hat H|\psi_n\rangle$};
    \node [block, right of=apply, node distance=2.5cm] (m_loop_enter) {$m:=1$};
    \node [decision, right of=m_loop_enter, node distance=2.0cm] (m_loop_cond) {$m=N$?};
    \node [decision, above of=m_loop_cond, node distance=2.2cm] (is_vanising) {$\langle\psi_m|\phi\rangle=0$?};
    \node [block, left of=is_vanising, node distance=2.2cm] (m_loop_inc) {$m:=m+1$};
    \node [block, above of=is_vanising, text width=10em, node distance=1.7cm] (find) {Find ${\cal H}_A: |\psi_n\rangle\in {\cal H}_A$\par Find ${\cal H}_B: |\psi_m\rangle\in {\cal H}_B$};
    \node [decision, above of=find, node distance=1.5cm] (is_same_subset) {${\cal H}_A = {\cal H}_B$?};
    \node [block, left of=is_same_subset, node distance=2.2cm] (merge) {Merge ${\cal H}_A$ and ${\cal H}_B$};
    \path [line] (input) -- (make_subsets);
    \path [line] (make_subsets) -- (n_loop_enter);
    \path [line] (n_loop_enter) -- (n_loop_cond);
    \path [line] (n_loop_cond) -- node {No} (apply);
    \path [line] (n_loop_cond) -- node {Yes} (result);
    \path [line] (apply) -- (m_loop_enter);
    \path [line] (m_loop_enter) -- (m_loop_cond);
    \path [line] (m_loop_cond) |- node {Yes} (n_loop_inc);
    \path [line] (n_loop_inc) -- (n_loop_cond);
    \path [line] (m_loop_cond.east) -| node [below] {No} +(0.5,1.5) |- (is_vanising.east);
    \path [line] (is_vanising) -- node [above] {Yes} (m_loop_inc);
    \path [line] (m_loop_inc) |- +(1,-1.2) -| (m_loop_cond);
    \path [line] (is_vanising) -- node {No} (find);
    \path [line] (find) -- (is_same_subset);
    \path [line] (is_same_subset) -- node {No} (merge);
    \path [line] (is_same_subset) |- node [above] {Yes} +(-3.3,1) |- (m_loop_inc);
    \path [line] (merge) -- (m_loop_inc);
\end{tikzpicture}
}
    \caption{(Color online) 
             Phase 1 of the automatic partitioning algorithm.
             A partition of the local Hilbert space satisfies condition
             (1), as described in the text, at the end of this phase.
             \label{fig:autopartition_phase1}}
\end{figure}

\underline{Phase 1} [Fig.~\ref{fig:autopartition_phase1}]

To start, one creates a data structure, which stores information
about the way $N$ basis states are partitioned into a number of subsets.
Initially each basis state resides alone in its own subset.
In the main loop of the algorithm, the Hamiltonian is sequentially
applied to each basis state (initial state). The application gives a linear
combination of the basis states with only a few non-zero coefficients,
since $\hat H$ is usually sparse.
The algorithm iterates in an inner loop over all basis vectors with non-zero
coefficients (final states).
If the initial and final state reside in different subsets, these subsets are
merged together. Once the main loop is over, the partition of the basis is
done. Two basis vectors are guaranteed to be in different subsets if they
cannot be reached from each other by application of $\hat H$ any number of times.
All matrix elements of $\hat H$ are calculated along the way and are
stored for later use.

\begin{figure}[H]
    \framebox[\textwidth]{
\begin{tikzpicture}[auto]
font issue=\footnotesize

\begin{scope} 
\fill[cyan!40!white] (-0.49\textwidth,0cm) rectangle (0.49\textwidth,-8cm) node[fitting node] (background) {};

\node [terminal,text width=7em] at ($(background.north east)-(1.8cm,0.7cm)$) (input)
    {Subsets $\{{\cal H}_A\}$,\\ operators $c^\ndag_\alpha, c_\alpha$};
\node [block, text width=12em, left of=input, node distance=8cm] (create_connections)
    {Create connection lists $\overline{M}$ and $\underline{M}$ (${\cal H}_A\to {\cal H}_B$)};
\node[block,below of=input, text width=6em, node distance=6.5cm] (apply_c) {$|\phi\rangle = c_\alpha|\psi_n\rangle$};
\node[decision,above of=apply_c,node distance=1.5cm] (is_vanishing_for_c) {$|\phi\rangle=0$?};
\node[block,above of=is_vanishing_for_c,text width=9em,node distance=1.5cm] (find_S_B_for_c) {Find ${\cal H}_B: |\phi\rangle\in {\cal H}_B$};
\node[block,above of=find_S_B_for_c,text width=9em,node distance=1.5cm] (add_to_M_down) {Add $(A,B)$ to $\underline{M}$};
\node[block,left of=apply_c,text width=9em,node distance=3.5cm] (find_S_A) {Find ${\cal H}_A : |\psi_n\rangle\in {\cal H}_A$};
\node[block,left of=is_vanishing_for_c,text width=6em,node distance=3.5cm] (apply_cd) {$|\phi\rangle = c^\dagger_\alpha|\psi_n\rangle$};
\node[decision,left of=find_S_B_for_c,node distance=3.5cm] (is_vanishing_for_cd) {$|\phi\rangle=0$?};
\node[block,left of=add_to_M_down,text width=9em,node distance=3.5cm] (find_S_B_for_cd) {Find ${\cal H}_B: |\phi\rangle\in {\cal H}_B$};
\node[block,above of=find_S_B_for_cd,text width=9em,node distance=1.5cm] (add_to_M_up) {Add $(A,B)$ to $\overline{M}$};
\node[block,left of=is_vanishing_for_cd,node distance=3.5cm] (loop_inc) {$n:=n+1$};
\node[decision,left of=find_S_A,node distance=3.5cm] (loop_cond) {$n=N$?};
\node[block,left of=loop_cond,text width=4em,node distance=3.5cm] (loop_enter) {$n:=1$};

\node[draw,rectangle] (label) at ($(background.north west)+(2.5cm,-2cm)$) {Fill $\overline{M}$ and $\underline{M}$};

\node[decision,below of=loop_cond,inner sep=-0.3em,node distance=1.85cm] (is_M_up_empty) {Is $\overline{M}$ empty?};
\node[terminal,left of=is_M_up_empty,node distance=3cm] (result) {Result: $\{{\cal H}_A\}$};
\node[block,right of=is_M_up_empty,text width=8em,node distance=3cm] (choose_root) {Choose some $(A,B)$ from $\overline{M}$};
\node[block,right of=choose_root,text width=10em,node distance=4cm] (call_zigzag_up) {Call \texttt{zigzag\_up($A$,$A$)}};

\path[line] (input) -- (create_connections);
\path[line] (create_connections) -| (loop_enter);
\path[line] (loop_enter) -- (loop_cond);
\path[line] (loop_cond) -- node {No} (find_S_A);
\path[line] (find_S_A) -- (apply_c);
\path[line] (apply_c) -- (is_vanishing_for_c);
\path[line] (is_vanishing_for_c) -- node[right] {No} (find_S_B_for_c);
\path[line] (is_vanishing_for_c.west) -| node[above,pos=0.3] {Yes} ($ (is_vanishing_for_c.west)!0.25!(apply_cd.south) $)
                                                        |- ($ (apply_cd.south) - (0,0.5cm)$) -| (apply_cd.south);
\path[line] (find_S_B_for_c) -- (add_to_M_down);
\path[line] (add_to_M_down.west)  -| ($ (add_to_M_down.west)!0.3!(apply_cd.east) $) |- (apply_cd.east);
\path[line] (apply_cd) -- (is_vanishing_for_cd);
\path[line] (is_vanishing_for_cd) -- node[right] {No} (find_S_B_for_cd);
\path[line] (is_vanishing_for_cd) -- node[above] {Yes} (loop_inc);
\path[line] (find_S_B_for_cd) -- (add_to_M_up);
\path[line] (add_to_M_up) -| (loop_inc);
\path[line] (loop_inc) -- (loop_cond);
\path[line] (loop_cond) -- node[right] {Yes} (is_M_up_empty);

\path[line] (is_M_up_empty) -- node[above] {Yes} (result);
\path[line] (is_M_up_empty) -- node[above] {No} (choose_root);
\path[line] (choose_root) -- (call_zigzag_up);
\path[line] (call_zigzag_up.south) |- (is_M_up_empty.south);
\end{scope}

\begin{scope} 
\fill[cyan!40!white] (-0.40\textwidth,-10cm) rectangle (-0.1cm,-15cm) node[fitting node] (background) {};

\node[draw,rectangle,anchor=north west] at (background.north west) (label) {\texttt{zigzag\_up($a$,$a_P$)}};
\node[draw,rectangle,anchor=south west] at (background.south west) (exit) {Exit};
\node[decision,below of=label,text width=5em,inner sep=0] (exists) {$\exists b:$\\$(a,b)\in\overline{M}$?};
\node[block,anchor=south east,text width=10em] at (background.south east) (remove) {Remove $(a,b)$ from $\overline{M}$};
\node[decision] at (remove |- exists) (is_parent) {$b=a_P$?};
\node[block,above of=is_parent, text width=10em, node distance=1.4cm] (merge) {Merge ${\cal H}_b$ and ${\cal H}_B$};
\node[block,above of=merge,text width=10em] (call_zigzag_down) {Call \texttt{zigzag\_down($b$,$a$)}};

\path[line] (label) -- (exists);
\path[line] (exists.west) |- ($ (exit.north)!0.5!(exists.west) $) -| node[pos=0.7] {No} (exit.north);
\path[line] (exists) |- node[pos=0.7] {Yes} (remove);
\path[line] (remove) -- (is_parent);
\path[line] (is_parent) -- node {Yes} (exists);
\path[line] (is_parent) -- node[right] {No} (merge);
\path[line] (merge) -- (call_zigzag_down);
\path[line] (call_zigzag_down) -| (exists);
\end{scope}

\begin{scope} 
\fill[cyan!40!white] (0.1cm,-10cm) rectangle (0.40\textwidth,-15cm) node[fitting node] (background) {};

\node[draw,rectangle,anchor=north west] at (0.1cm,-10cm) (label) {\texttt{zigzag\_down($b$,$b_P$)}};
\node[draw,rectangle,anchor=south west] at (0.1cm,-15cm) (exit) {Exit};
\node[decision,below of=label,text width=5em,inner sep=0] (exists) {$\exists a:$\\$(b,a)\in\underline{M}$?};
\node[block,anchor=south east,text width=10em] at (background.south east) (remove) {Remove $(b,a)$ from $\underline{M}$};
\node[decision] at (remove |- exists) (is_parent) {$a=b_P$?};
\node[block,above of=is_parent, text width=10em, node distance=1.4cm] (merge) {Merge ${\cal H}_a$ and ${\cal H}_A$};
\node[block,above of=merge,text width=10em] (call_zigzag_down) {Call \texttt{zigzag\_up($a$,$b$)}};

\path[line] (label) -- (exists);
\path[line] (exists.west) |- ($ (exit.north)!0.5!(exists.west) $) -| node[pos=0.7] {No} (exit.north);
\path[line] (exists) |- node[pos=0.7] {Yes} (remove);
\path[line] (remove) -- (is_parent);
\path[line] (is_parent) -- node {Yes} (exists);
\path[line] (is_parent) -- node[right] {No} (merge);
\path[line] (merge) -- (call_zigzag_down);
\path[line] (call_zigzag_down) -| (exists);
\end{scope}

\end{tikzpicture}
}
    \caption{(Color online) 
             Phase 2 of the automatic partitioning algorithm.
             A partition of the local Hilbert space satisfies condition
             (2), as described in the text, after this phase has been applied to all
             operator pairs $c^\dag_\alpha,c_\alpha$.
             \label{fig:autopartition_phase2}}
\end{figure}
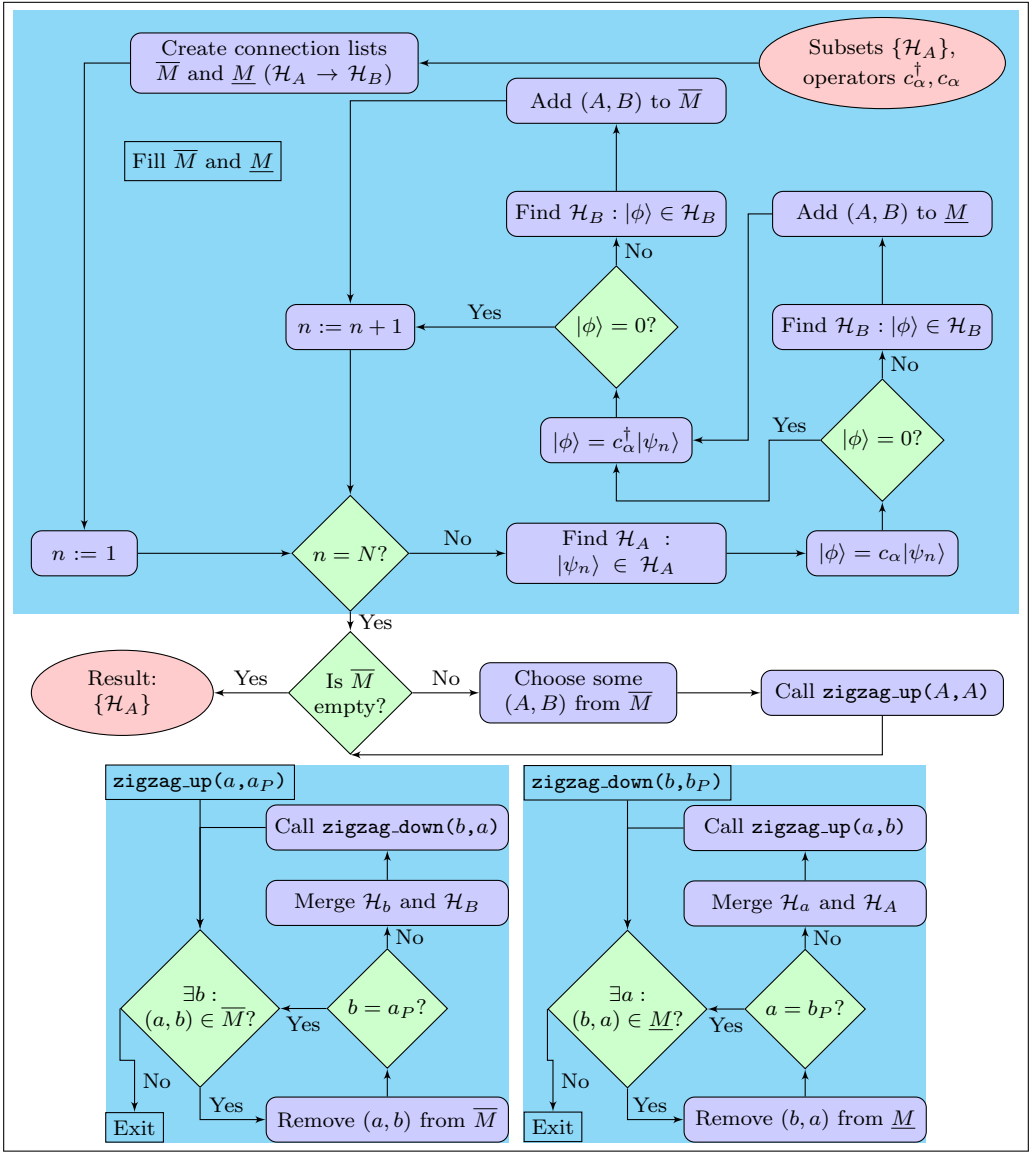

\underline{Phase 2} [Fig.~\ref{fig:autopartition_phase2}]

During this phase some subsets are additionally merged to satisfy condition
(2). This part of the algorithm is executed in turn for all $c^\ndag_\alpha,
c_\alpha$ pairs, and for each pair it proceeds as follows.

Two lists of subspace-to-subspace connections are first generated:
$\overline{M} = \{{\cal H}_A\to {\cal H}_B\}$ for $c^\ndag_\alpha$, and 
$\underline{M} = \{{\cal H}_A\to {\cal H}_B\}$ for $c_\alpha$.
These lists are created by direct applications of the corresponding operators
to all Fock states to determine, in which subspaces the resulting wave function
has nonzero amplitudes.

Then, the algorithm recursively iterates over all connections in $\overline{M}$
and $\underline{M}$, and merges some subspaces following a special `zigzag'
visiting pattern. Let us consider a tree-like structure with subspaces being
the nodes of the tree, and connections being the edges. We are interested in a
tree generated by a sequence of operators $c^\ndag_\alpha, c_\alpha
c^\ndag_\alpha, c^\ndag_\alpha c_\alpha c^\ndag_\alpha,\ldots \,$ from a randomly
chosen root subspace ${\cal H}_R$. The recursive procedure starts at the root and
traverses the tree in the depth-first order. 
Simultaneously it perform two actions:
\begin{itemize}
    \item Removes visited connections from $\overline{M}$ (edges from an odd
          level nodes), or from $\underline{M}$ (edges from an even level nodes). 
    \item Merges all odd level subspaces with the root subspace and all even
          level subspaces with the second level subspace.
\end{itemize}
If the tree has been fully traversed, but $\overline{M}$ is still not empty, another
connection is picked from $\overline{M}$. This connection serves as a root$\to$first
level branch of the next tree to be traversed.

The proposed `zigzag' traversal procedure squeezes every tree to a pair of
nodes connected by one edge. As every connection is visited only once, this is
a computationally economic way to ensure fulfillment of condition (2).

An important remark should be made about the data structure which maintains the
partition of ${\cal H}$. We use a \textit{disjoint set data structure} designed
especially for quick {\bf find\_set}, {\bf union\_set}, and {\bf make\_set}
operations \cite{tarjan,boost_dsets}.
Thanks to two special techniques called `union by rank' and
`path compression', the amortized time per operation is only ${\cal O}(\alpha(n))$.
Here $n$ is the total number of {\bf find\_set}, {\bf union\_set}, and {\bf make\_set}
operations, and $\alpha(n)$ is the extremely slowly-growing
inverse Ackermann function, with $\alpha(n) < 5$ for any practical value of $n$.

A direct comparison of the two partitioning schemes shows that they discover equal
numbers of subspaces in the cases of the 2-,3-,4-,5-,6- and 7-orbital
Hubbard-Kanamori Hamiltonians (provided the quantum numbers of Ref.
\cite{parragh_conserved_2012} are used in addition to $\hat N$ and $\hat S_z$).
For the rotationally-invariant Slater Hamiltonians, however, the automatic
partition algorithm is advantageous, as shown in Table~\ref{tab:autovsqn}.

\begin{table}[htpb]
\centering
\begin{tabular}{lcc}
\hline
Model & \#subspaces, QN & \#subspaces, Automatic\\
\hline
\hline
5 orbitals ($c^\dag_\alpha c_\alpha$ in spherical basis) & 36 & 276 \\ 
7 orbitals ($c^\dag_\alpha c_\alpha$ in spherical basis) & 64 & 960 \\ 
5 orbitals ($c^\dag_\alpha c_\alpha$ in cubic basis)     & 36 & 132 \\
7 orbitals ($c^\dag_\alpha c_\alpha$ in cubic basis)     & 64 & 244 \\ 
\hline
\end{tabular}
  \caption{Comparison of the number of subspaces resulting from the use of
           quantum numbers (QN) and the partitioning algorithm (Automatic) presented
           in this paper for rotationally-invariant Slater Hamiltonians.
    \label{tab:autovsqn}}
\end{table}

\section{Efficient trace calculation using a balanced tree and truncation}
\label{sec:tree}

\begin{figure}[H]
  \centering
    \includegraphics[width=0.85\textwidth]{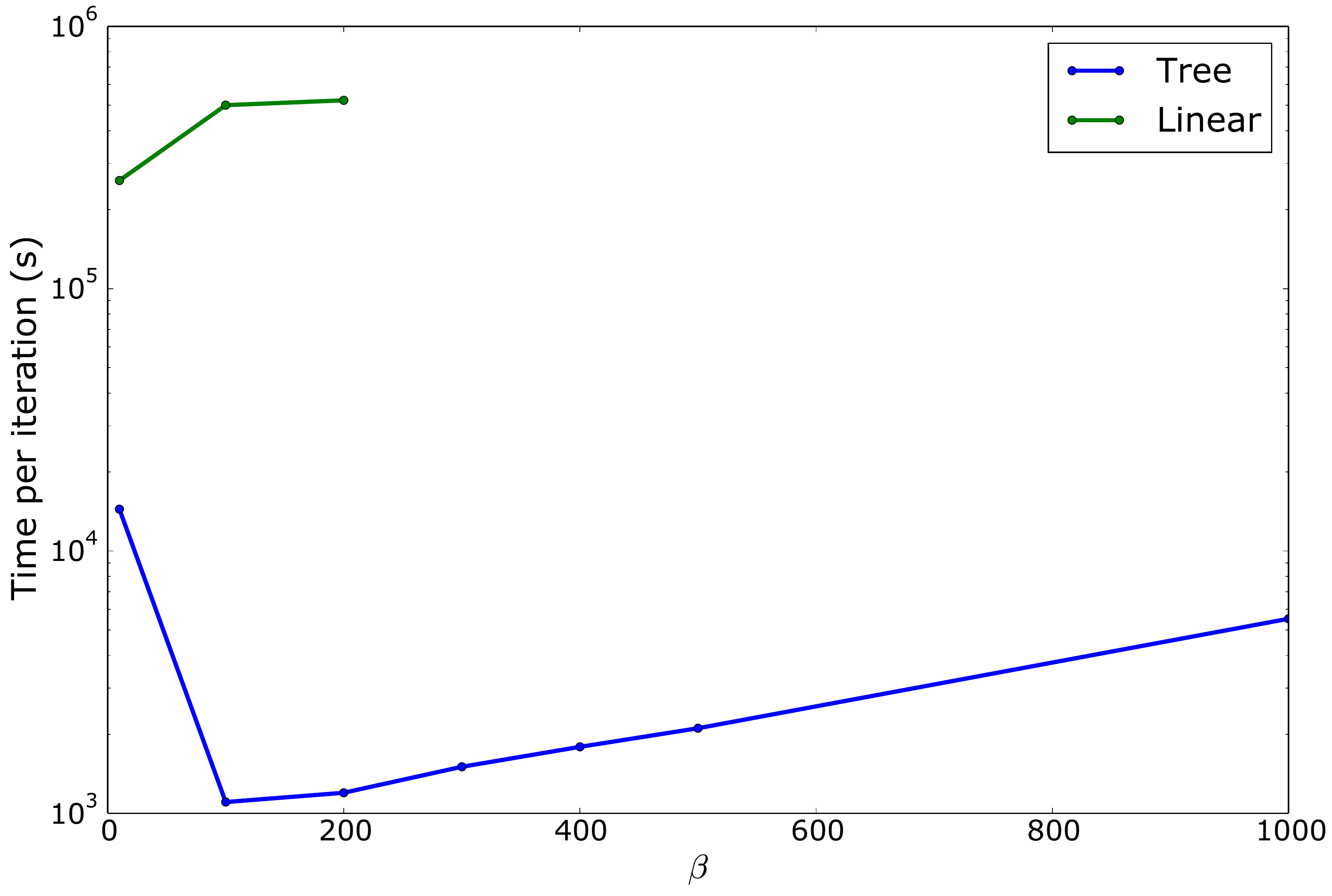}
    \caption{(Color online)
             Scaling of computational time with the inverse temperature $\beta$ 
             for a system with three electrons in five bands in a Bethe lattice 
             with a fully rotationally-invariant Hamiltonian for two algorithms:
             a) the \emph{linear} case in which the trace is recomputed from
             the full linear chain of operator matrices and 
	     b) the \emph{tree} algorithm proposed here, in which the number
	     of matrix products necessary is reduced.  The overall scaling
             depends both on the scaling of the tree (${\cal O}({\rm log}_2 \beta)$)
             and that of the determinants (${\cal O}(\beta^3)$).
             The same number of Monte Carlo steps is used in all calculations.
             \label{fig:3ein5Bands}}
\end{figure}

Here we discuss the optimisation of the calculation of the atomic trace using a
tree structure. We emphasize that the use of such a tree does not change the
Markov chain, but is simply a method to reduce the computational cost of the
trace computation. The principal idea behind the CT-HYB algorithm presented
here is to use a balanced red-black tree to describe the configurations in the
Markov chain. The use of a tree to store a configuration and partial products
necessary in the trace evaluation reduces the number of matrix products that
need to be performed when inserting or removing a collection of one or more
operator pairs from the configuration. Hence, the cost of this limiting step of
the algorithm is significantly reduced as compared to the na\"ive `linear'
algorithm where the configuration is a linear chain of operators and all the
matrix products are explicitly evaluated for each configuration. The gain in
time per iteration between the tree and linear methods of computing the trace
for a five-band system with three electrons and an rotationally-invariant
interaction is shown in Fig.~\ref{fig:3ein5Bands}. Such a representation was
first proposed by Gull \cite{gull_continuous-time_2008}. Similar
implementations based on skip lists have already appeared
\cite{semon_lazy_2014,huang_iqist_2015} and are shown to be successful despite
not mathematically guaranteeing logarithmic scaling.

Further efficiency gains are also made by using bound properties of the
trace to quickly reject proposed moves \cite{yee_towards_2012}.
Additionally, by summing over the blocks that contribute to the trace in order
of increasing importance (i.e., increasing energies), one is able to truncate
the trace evaluation once further contributions are smaller than machine
precision~\cite{semon_lazy_2014}.

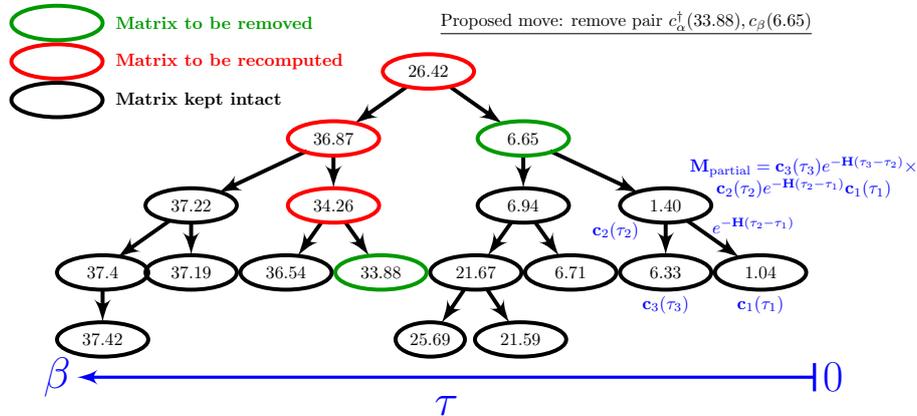
\begin{figure}[H]
  \begin{center}
 \begin{adjustbox}{width=0.90\textwidth}
\begin{tikzpicture}[>=latex',line join=bevel,scale=0.23,every node/.style={scale=0.45}]
  \pgfsetlinewidth{1bp}
  \pgfsetcolor{black}
  \draw [->] (315.85bp,148.71bp) .. controls (321.98bp,139.99bp) and (329.50bp,129.4bp)  .. (342.11bp,111.34bp);
  \draw [->] (540.34bp,227.08bp) .. controls (565.21bp,215.28bp) and (601.07bp,198.27bp)  .. (637.52bp,180.98bp);
  \draw [->] (273.66bp,227.08bp) .. controls (248.79bp,215.28bp) and (212.93bp,198.27bp)  .. (176.48bp,180.98bp);
  \draw [->] (429.82bp,298.76bp) .. controls (444.45bp,288.35bp) and (463.64bp,274.7bp)  .. (488.23bp,257.2bp);
  \draw [->] (471.11bp,74.708bp) .. controls (476.88bp,65.991bp) and (483.9bp,55.395bp)  .. (495.86bp,37.338bp);
  \draw [->] (446.64bp,74.708bp) .. controls (440.71bp,65.926bp) and (433.48bp,55.236bp)  .. (421.22bp,37.082bp);
  \draw [->] (124.18bp,150.76bp) .. controls (109.55bp,140.35bp) and (90.359bp,126.7bp)  .. (65.77bp,109.2bp);
  \draw [->] (511bp,221.94bp) .. controls (511bp,214.15bp) and (511bp,204.95bp)  .. (511bp,186.25bp);
  \draw [->] (523.85bp,148.71bp) .. controls (529.98bp,139.99bp) and (537.50bp,129.4bp)  .. (550.11bp,111.34bp);
  \draw [->] (50bp,73.950bp) .. controls (50bp,66.149bp) and (50bp,56.954bp)  .. (50bp,38.249bp);
  \pgfsetcolor{black}
  \draw [->] (384.18bp,298.76bp) .. controls (369.55bp,288.35bp) and (350.36bp,274.7bp)  .. (325.77bp,257.2bp);
  \draw [->] (689.82bp,150.76bp) .. controls (704.45bp,140.35bp) and (723.64bp,126.7bp) .. (748.23bp,109.2bp) node[above] at +(15pt,20pt) [blue] {$e^{-\mathbf{H}(\tau_2-\tau_1)}$}; 
  \draw [->] (303bp,221.94bp) .. controls (303bp,214.15bp) and (303bp,204.95bp)  .. (303bp,186.25bp);
  \draw [->] (498.15bp,148.71bp) .. controls (492.02bp,139.99bp) and (484.57bp,129.4bp)  .. (471.89bp,111.34bp);
  \pgfsetcolor{black}
  \draw [->] (147bp,147.94bp) .. controls (147bp,140.15bp) and (147bp,130.95bp)  .. (147bp,112.25bp);
  \draw [->] (290.15bp,148.71bp) .. controls (284.02bp,139.99bp) and (276.57bp,129.4bp)  .. (263.89bp,111.34bp);
  \draw [->] (667bp,147.94bp) .. controls (667bp,140.15bp) and (667bp,130.95bp)  .. (667bp,112.25bp);
\begin{scope}
  \definecolor{strokecol}{rgb}{0.0,0.0,0.0};
  \pgfsetstrokecolor{strokecol}
  \draw (409bp,19bp) ellipse (38bp and 19bp);
  \draw (409bp,19bp) node {$25.69$};
\end{scope}
\begin{scope}
  \definecolor{strokecol}{rgb}{0.0,0.6,0.0};
  \pgfsetstrokecolor{strokecol}
  \draw (511bp,241bp) ellipse (50bp and 19bp);
  \definecolor{strokecol}{rgb}{0.0,0.0,0.0};
  \pgfsetstrokecolor{strokecol}
  \draw (511bp,241bp) node {$6.65$};
\end{scope}
\begin{scope}
  \definecolor{strokecol}{rgb}{1.0,0.0,0.0};
  \pgfsetstrokecolor{strokecol}
  \draw (303bp,241bp) ellipse (50bp and 19bp);
  \definecolor{strokecol}{rgb}{0.0,0.0,0.0};
  \pgfsetstrokecolor{strokecol}
  \draw (303bp,241bp) node {$36.87$};
\end{scope}
\begin{scope}
  \definecolor{strokecol}{rgb}{0.0,0.0,0.0};
  \pgfsetstrokecolor{strokecol}
  \draw (147bp,167bp) ellipse (50bp and 19bp);
  \draw (147bp,167bp) node {$37.22$};
\end{scope}
\begin{scope}
  \definecolor{strokecol}{rgb}{0.0,0.0,0.0};
  \pgfsetstrokecolor{strokecol}
  \draw (667bp,93bp) ellipse (50bp and 19bp);
  \draw (667bp,93bp) node (c3) {$6.33$};
  \node[below=1pt of c3] [blue] {$\mathbf{c}_3(\tau_3)$};
\end{scope}
\begin{scope}
  \definecolor{strokecol}{rgb}{0.0,0.0,0.0};
  \pgfsetstrokecolor{strokecol}
  \draw (511bp,167bp) ellipse (50bp and 19bp);
  \draw (511bp,167bp) node {$6.94$};
\end{scope}
\begin{scope}
  \definecolor{strokecol}{rgb}{0.0,0.0,0.0};
  \pgfsetstrokecolor{strokecol}
  \draw (50bp,19bp) ellipse (50bp and 19bp);
  \draw (50bp,19bp) node {$37.42$};
\end{scope}
\begin{scope}
  \definecolor{strokecol}{rgb}{0.0,0.0,0.0};
  \pgfsetstrokecolor{strokecol}
  \draw (508bp,19bp) ellipse (50bp and 19bp);
  \draw (508bp,19bp) node {$21.59$};
\end{scope}
\begin{scope}
  \definecolor{strokecol}{rgb}{0.0,0.0,0.0};
  \pgfsetstrokecolor{strokecol}
  \draw (147bp,93bp) ellipse (50bp and 19bp);
  \draw (147bp,93bp) node {$37.19$};
\end{scope}
\begin{scope}
  \definecolor{strokecol}{rgb}{0.0,0.0,0.0};
  \pgfsetstrokecolor{strokecol}
  \draw (563bp,93bp) ellipse (50bp and 19bp);
  \draw (563bp,93bp) node {$6.71$};
\end{scope}
\begin{scope}
  \definecolor{strokecol}{rgb}{0.0,0.0,0.0};
  \pgfsetstrokecolor{strokecol}
  \draw (50bp,93bp) ellipse (50bp and 19bp);
  \draw (50bp,93bp) node {$37.4$};
\end{scope}
\begin{scope}
  \definecolor{strokecol}{rgb}{1.0,0.0,0.0};
  \pgfsetstrokecolor{strokecol}
  \draw (407bp,315bp) ellipse (50bp and 19bp);
  \definecolor{strokecol}{rgb}{0.0,0.0,0.0};
  \pgfsetstrokecolor{strokecol}
  \draw (407bp,315bp) node {$26.42$};
\end{scope}
\begin{scope}
  \definecolor{strokecol}{rgb}{0.0,0.0,0.0};
  \pgfsetstrokecolor{strokecol}
  \draw (667bp,167bp) ellipse (50bp and 19bp);
  \draw (667bp,167bp) node (c2) {$1.40$};
  \node[below left=-1pt of c2] [blue] {$\mathbf{c}_2(\tau_2)$};
  \node[above right=-5pt of c2] [blue] {$\begin{array}{c}
                                  \mathbf{M}_\mathrm{partial} = \mathbf{c}_3(\tau_3) e^{-\mathbf{H}(\tau_3-\tau_2)}\times\\
                                  \mathbf{c}_2(\tau_2) e^{-\mathbf{H}(\tau_2-\tau_1)} \mathbf{c}_1(\tau_1)\end{array}$};
\end{scope}
\begin{scope}
  \definecolor{strokecol}{rgb}{0.0,0.0,0.0};
  \pgfsetstrokecolor{strokecol}
  \draw (459bp,93bp) ellipse (50bp and 19bp);
  \draw (459bp,93bp) node {$21.67$};
\end{scope}
\begin{scope}
  \definecolor{strokecol}{rgb}{0.0,0.0,0.0};
  \pgfsetstrokecolor{strokecol}
  \draw (251bp,93bp) ellipse (50bp and 19bp);
  \draw (251bp,93bp) node {$36.54$};
\end{scope}
\begin{scope}
  \definecolor{strokecol}{rgb}{1.0,0.0,0.0};
  \pgfsetstrokecolor{strokecol}
  \draw (303bp,167bp) ellipse (50bp and 19bp);
  \definecolor{strokecol}{rgb}{0.0,0.0,0.0};
  \pgfsetstrokecolor{strokecol}
  \draw (303bp,167bp) node {$34.26$};
\end{scope}
\begin{scope}
  \definecolor{strokecol}{rgb}{0.0,0.6,0.0};
  \pgfsetstrokecolor{strokecol}
  \draw (355bp,93bp) ellipse (50bp and 19bp);
  \definecolor{strokecol}{rgb}{0.0,0.0,0.0};
  \pgfsetstrokecolor{strokecol}
  \draw (355bp,93bp) node {$33.88$};
\end{scope}
\begin{scope}
  \definecolor{strokecol}{rgb}{0.0,0.0,0.0};
  \pgfsetstrokecolor{strokecol}
  \draw (771bp,93bp) ellipse (50bp and 19bp);
  \draw (771bp,93bp) node (c1) {$1.04$};
  \node[below=1pt of c1] [blue] {$\mathbf{c}_1(\tau_1)$};
\end{scope}

\begin{scope}
   \node (tau_0) at (30,-0.8) [blue] {\Huge $0$};
   \node (tau_beta) at (0,-0.8) [blue] {\Huge $\beta$};
   \draw[|->,thick,blue] (tau_0) -- (tau_beta) node [pos=0.5,below=3pt] {\Huge $\tau$};
\end{scope}

\begin{scope}
    \draw[darkgreen] (0,13) ellipse (50bp and 19bp) node[anchor=west] at (2,13) {{\normalsize\bf Matrix to be removed}};
    \draw[red] (0,11.5) ellipse (50bp and 19bp) node[anchor=west] at (2,11.5) {{\normalsize\bf Matrix to be recomputed}};
    \draw[black] (0,10) ellipse (50bp and 19bp) node[anchor=west] at (2,10) {{\normalsize\bf Matrix kept intact}};
    \node at (22,13) {{\normalsize\underline{Proposed move: remove pair $c^\dag_{\alpha}(33.88), c_{\beta}(6.65)$}}};
\end{scope}

\end{tikzpicture}
 \end{adjustbox}
 \end{center}
 \caption{
  \label{fig:tree}
  (Color online)
  The representation of a configuration by a tree. Each node represents an
  operator at the time that determines the node's key. The nodes are sorted
  according to decreasing time to reflect the time-ordering of the trace. When
  the removal of the pair of operators at $\tau=33.88$ and $\tau=6.65$
  (highlighted in green) is proposed, only the contents (specifically the partial
  product of matrices) of the nodes between these nodes up to the root node at
  $\tau=26.42$ (highlighted in red) need to be updated. Specifically, the matrix
  stored at each node contains the product of the partial product on each of its
  subtree nodes, the operator of the node itself and the respective time
  evolution operators as shown in the figure.
 }
\end{figure}

Each configuration in the Markov chain is described by a tree with operators as
nodes consisting of a key-value pair. The key, based on which the tree is
sorted, is given by the imaginary time $\tau$ of the operator. The value
consists of the operator and the matrix product of the subtree (using a block
structure). The tree for a configration is depicted in Fig.~\ref{fig:tree}. 

We implement the left-leaning red-black tree (LLRBT), as explained in 
Refs.~\citealp{sedgewick_left-leaning_2008,sedgewick_algorithms_coursera}. Our C++
implementation of the LLRBT is adapted from the Java code given in
Ref.~\citealp{sedgewick_algorithms_2011}.

In particular, the basic implementation of a LLRBT was adapted to minimise the
rebalancing of the tree for a Metropolis Monte Carlo algorithm in which
attempted moves are not always accepted. At each proposed Monte Carlo step, we
add and/or remove nodes in the tree and determine the corresponding trace {\it
without} rebalancing the tree. If the proposed move is accepted, the tree is
balanced.

As seen in Fig.~\ref{fig:tree}, the evaluation of a single insertion/removal
requires at most ${\rm log}_{2}(K)$ matrix products (i.e., the height of the
tree from the root node to the deepest nodes) where $K$ is the order of the
configuration. The cost of the trace computation scales logarithmically with
the perturbation order of the configuration rather than linearly in the linear
algorithm.

\begin{figure}[H]
\begin{center}
\includegraphics[width=0.85\textwidth]{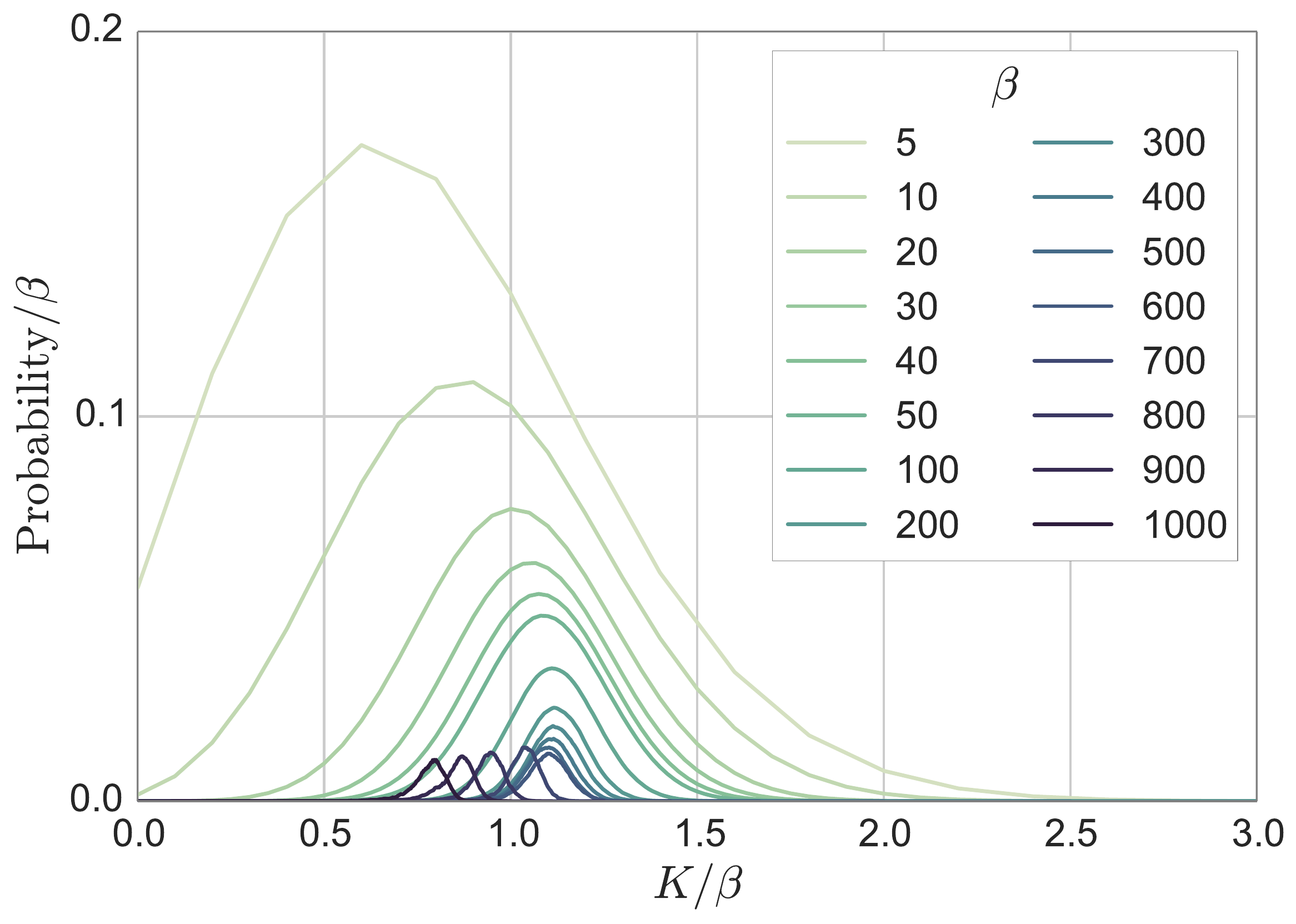} 
\end{center}
\caption{
 \label{fig:pert_order} 
 (Color online) 
 The perturbation order $K$ increases approximately linearly with $\beta$.
}
\end{figure}

The average perturbation order of the partition function increases
approximately linearly with the inverse temperature $\beta$ as shown in
Ref.~\citealp{haule_quantum_2007} and in Fig.~\ref{fig:pert_order}. At low
$\beta$, the cost of computing the trace dominates over that of the
determinant. For sufficiently large $\beta$, we expect that the computation of
the determinants, which scales as $K^3$, to dominate in each Monte Carlo step.

\begin{figure}[H]
\begin{center}
 \includegraphics[width=0.85\textwidth]{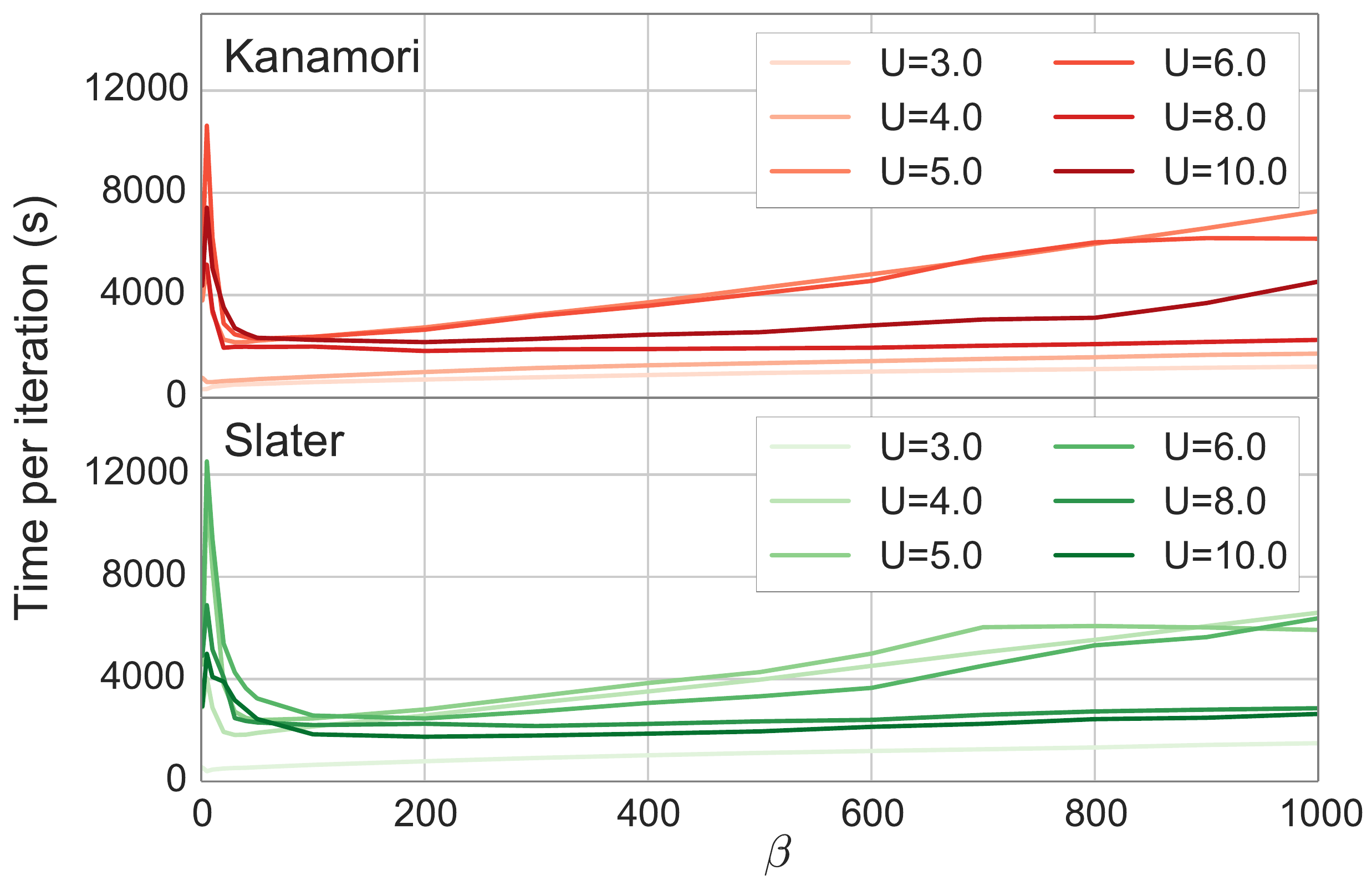} 
\end{center}
\caption{
 \label{fig:scaling} 
 (Color online) 
 The scaling of the cost of solving the impurity problem for Kanamori and fully
 rotationally-invariant Slater Hamiltonians with the inverse temperature $\beta$.
 The same number of Monte Carlo steps is used in all calculations.
}
\end{figure}

Hence, the tree allows us to reach much lower temperatures than was possible
with the na\"ive linear algorithm for a fixed number of MC steps. A rough
estimate of the scaling of the cost of five band calculations using both
approximate Kanamori and fully rotationally-invariant Slater parametrisation
of the interactions with $\beta$ is shown in
Fig.~\ref{fig:scaling}. 

\section{Four-operator moves and ergodicity considerations}
\label{sec:doublemoves}

\begin{figure}[H]
\begin{center}
 \includegraphics[width=0.85\textwidth]{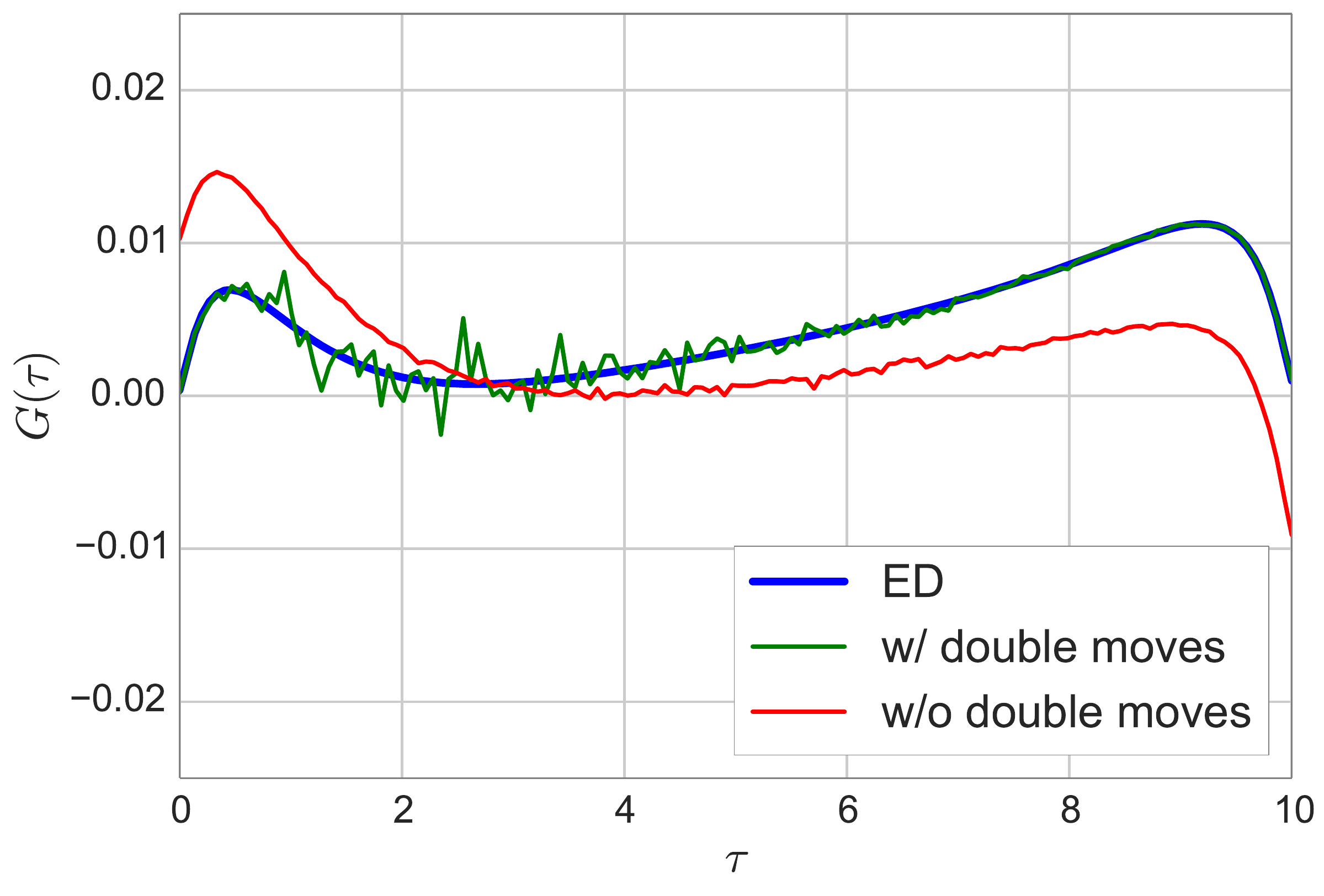} 
\end{center}
\caption{
 \label{fig:move_double} 
 (Color online) 
 The off-diagonal component of $G(\tau)$ for a two-band Kanamori Hamiltonian
 computed using \py{CTHYB} without double moves (red) differs from the exact
 diagonalisation \cite{antipov_pomerol_2015} result (blue). Inclusion of double
 moves (green) corrects this error. The higher level of noise in the result with
 double moves reflects the low acceptance rate of this move.
}
\end{figure}

Introduction of more complex moves beyond the commonly used insertion/removal
of a single pair of operators were shown to be important in symmetry-broken
cases \cite{semon_ergodicity_2014}. We have implemented Monte Carlo moves in
which four operators are simulataneously inserted/removed, and moreover, we
show that such moves are crucial for proper sampling of the configuration space
even in cases without symmetry-breaking. One example is a two-band Kanamori
model with off-diagonal components in the hybridization function. In
Fig.~\ref{fig:move_double}, we show the results of the calculation without
double moves, with double moves and using exact diagonalisation
\cite{antipov_pomerol_2015}. When double moves are not used, the results are
clearly incorrect. The use of double moves is necessary to arrive at the
correct result.

The ergodicity problem for such a multi-orbital Kanamori Hamiltonian in the
absence of four-operator moves can be understood. Let us assume that the
hybridization function is spin-diagonal but has orbital off-diagonal
components. Configurations such as $c^\dag_{1 \up} c^\dag_{2 \dn} c_{1 \dn}
c_{2 \up}$ yield a non-zero contribution to the trace for the two-particle
sector of eigenstates, and more precisely, for the Hund's singlet and triplet.
However, they can never be reached when making only pair insertions. Insertion
of the $c^\dag_{1 \up}$, $c_{1 \dn}$ and $c^\dag_{2 \dn}$, $c_{2 \up}$ pairs is
never proposed as the hybridization function is spin-diagonal. On the other
hand, $c^\dag_{1 \up}$, $c_{2 \up}$ and $c^\dag_{2 \dn}$, $c_{1 \dn}$ lead to a
structural cancellation for all eigenstates due to conservation of the
total number of operators on each orbital {\it modulo} 2. We have an obvious
ergodicity problem and the Monte Carlo chain cannot reach such a configuration.
On the other hand, it is clear that this configuration can be reached by a
four-operator insertion.

Here we discuss a concrete example of a configuration for which double moves
are necessary in the Kanamori case. The existence of ergodicity problems for
generic Hamiltonians in such algorithms is still an open issue. For example, it
has not yet been proven that four-operator insertions completely avoid
ergodicity problems in a five-band rotationally-invariant Slater Hamiltonian.
This is an important question that requires further investigation.

\section{Getting started}
\label{sec:starting}

\subsection{Obtaining \py{CTHYB}}

The \py{CTHYB} source code is available publicly and can be obtained by cloning
the repository on the GitHub website at
\href{https://github.com/TRIQS/cthyb}{https://github.com/TRIQS/cthyb}. As the
TRIQS project is continuously improving, we recommend that users always obtain
TRIQS and its applications, including \py{CTHYB}, from GitHub. Fixes to
possible issues are also applied to the GitHub source.

\subsection{Installation}
\label{sec:installation}

Installing \py{CTHYB} follows the same procedure as needed to install TRIQS. 
Here too, we use the \verb#cmake# tool to
configure, build and test the application. Assuming that TRIQS has been installed at 
\verb#/path/to/TRIQS/install/dir# (refer to the online documentation),
\py{CTHYB} is simply installed by issuing the following commands at the shell
prompt:
\begin{verbatim}
$ git clone https://github.com/TRIQS/cthyb.git src
$ mkdir build_cthyb && cd build_cthyb
$ cmake -DTRIQS_PATH=/path/to/TRIQS/install/dir ../src
$ make
$ make test
$ make install
\end{verbatim}
This will install \py{CTHYB} in the same location as TRIQS. Further
installation instructions can be found in the online documentation.

\subsection{Citation policy}

We kindly request that the present paper be cited in any published work using
the \py{CTHYB} solver. Furthermore, we ask that the TRIQS library on which the
solver presented here is based also be cited \cite{triqs_2015}. This helps the
\py{CTHYB} and TRIQS developers to better keep track of projects using the library
and provides them guidance for future developments.

\subsection{Contributing}
\py{CTHYB}, as an application of TRIQS, is an open source project and we
encourage feedback and contributions from the user community. Issues should be
reported exclusively via the GitHub website at
\href{https://github.com/TRIQS/cthyb/issues}{https://github.com/TRIQS/cthyb/issues}.
For contributions, we recommend to use the pull request system on the GitHub
website. Before any major contribution, we recommend coordination with the
main \py{CTHYB} developers.

\section{Summary}
\label{sec:summary}

We have presented the free software \py{TRIQS/CTHYB}, an implementation of the
continuous-time hybridization expansion quantum Monte Carlo impurity solver. In
addition to implementing the various improvements documented in the
literature, we have also presented and implemented a new algorithm
to divide the local Hilbert space, removing the need for the
user to explicitly provide the symmetry of the impurity Hamiltonian or its
quantum numbers. We also discussed a case where `double moves' in the quantum
Monte Carlo are required in order to obtain correct results even in the absence
of symmetry-breaking. Further developments (e.g., support for complex
Hamiltonians, other measures) are planned for a future release. 

\section{Acknowledgments}
\label{sec:acknowledgements}

We thank S.~Biermann, E.~Gull and P.~Werner for useful discussions,
M.~Aichhorn, B.~Amadon, T.~Ayral, P.~Delange, P.~Hansmann, M.~Harland,
L.~Pourovskii, W.~Rowe, M.~Zingl for their feedback on the code.  The TRIQS
project is supported by the ERC Grant No.~278472--\emph{MottMetals}.  I.~K.~
acknowledges support from Deutsche Forschungsgemeinschaft via Project SFB
668-A3. P.~S.~ acknowledges support from ERC Grant
No.~617196--\emph{CorrelMat}.

\bibliographystyle{elsarticle-num.bst}
\bibliography{main}

\end{document}